\documentclass[aps,prl,showpacs,amsmath,amssymb,longbibliography, notitlepage,superscriptaddress, twocolumn]{revtex4-1} %,twocolumn, pre or preprint
\usepackage[dvipsnames]{xcolor}
\usepackage{floatrow}
\usepackage{soul}
\usepackage{siunitx}
\soulregister\cite7
\soulregister\ref7

\usepackage{graphicx}
\graphicspath{{../img/}}
\usepackage{placeins}

\definecolor{darkblue}{rgb}{0,0,0.6}
\definecolor{darkred}{rgb}{0.6,0,0}
\usepackage[colorlinks=true,urlcolor=darkblue,citecolor=darkblue, linkcolor=darkred,hyperfootnotes=false]{hyperref}

\usepackage[color=yellow,]{todonotes}

\RequirePackage{lettrine} % For dropped capitals

\setul{}{1.2pt}

\newcommand{\srr}{\sigma_{rr}}
\newcommand{\stt}{\sigma_{\theta\theta}}

\begin{document}

\title{Sculpting liquids with ultrathin shells}

\author{Yousra Timounay}
%\email{ytimouna@syr.edu}
\affiliation{Department of Physics, Syracuse University, Syracuse, NY 13244}
\affiliation{BioInspired Syracuse: Institute for Material and Living Systems, Syracuse University, Syracuse, NY 13244}
\author{Alexander R. Hartwell}
\affiliation{Department of Physics, Syracuse University, Syracuse, NY 13244}
\author{Mengfei He}
\affiliation{Department of Physics, Syracuse University, Syracuse, NY 13244}
\affiliation{BioInspired Syracuse: Institute for Material and Living Systems, Syracuse University, Syracuse, NY 13244}
\author{D. Eric King}
\affiliation{Department of Physics, Syracuse University, Syracuse, NY 13244}
\author{Lindsay K. Murphy}
\affiliation{Department of Physics, Syracuse University, Syracuse, NY 13244}
%\author{Graham C. Leggat}
%\affiliation{Department of Physics, Syracuse University, Syracuse, NY 13244}
\author{Vincent D\'emery}
\email{vincent.demery@espci.psl.eu}
\affiliation{Gulliver UMR CNRS 7083, ESPCI Paris, Universit\'e PSL, (10 rue Vauquelin) 75005 Paris, France}
\affiliation{Univ Lyon, ENS de Lyon, Univ Claude Bernard Lyon 1, CNRS, Laboratoire de Physique, F-69342 Lyon, France}
\author{Joseph D. Paulsen}
\email{jdpaulse@syr.edu}
\affiliation{Department of Physics, Syracuse University, Syracuse, NY 13244}
\affiliation{BioInspired Syracuse: Institute for Material and Living Systems, Syracuse University, Syracuse, NY 13244}

\begin{abstract}
    %Thin elastic structures may be bent, twisted, or stretched by the surface tension of a liquid. Understanding these deformations is essential for the manufacture and control of fibers, sheets, and shells that reside on a liquid or may become wetted. 
    Thin elastic films can spontaneously attach to liquid interfaces, offering a platform for tailoring their physical, chemical, and optical properties. 
    Current understanding of the elastocapillarity of thin films is based primarily on studies of planar sheets. 
    We show that curved shells can be used to manipulate interfaces in qualitatively different ways. 
    We elucidate a regime where an ultrathin shell with vanishing bending rigidity imposes its own rest shape on a liquid surface, using experiment and theory. 
    %Conceptually, the shell may be thought of as ``inflating'' into its rest configuration by the pressure across the interface. 
    Conceptually, the pressure across the interface ``inflates'' the shell into its original shape. 
    The setup is amenable to optical applications as the shell is transparent, free of wrinkles, and may be manufactured over a range of curvatures. 
\end{abstract}

\maketitle

%\section{Introduction}

%Thin elastic films can spontaneously attach to fluid interfaces in a process driven by capillary forces
Capillary forces %are strong enough to 
can anchor a sufficiently thin elastic solid onto a fluid interface
\cite{Huang07,Gao08,Kumar20}. 
Such adsorbed films offer a means to control interfaces by modifying their shape~\cite{Paulsen15}, mechanics~\cite{Vella15,Ripp20}, or permeability~\cite{Kumar18}, or by providing a substrate for physical or chemical patterning~\cite{Reynolds19}. 
Crucial to such applications is an understanding of how geometric incompatibilities between a film and an interface are resolved \cite{Hure11,King12,Davidovitch19}. % also Hure12
%Thin elastomer films may undergo long-wavelength buckling when confined \cite{Py07,Pezzulla18,Bense20}; 
Here we focus on ultrathin ($\sim 100~\si{\nano m}$) polymer films that strongly resist in-plane stretching yet readily wrinkle, allowing them to conform to a wide range of surface topographies \cite{Paulsen16}. 
Such films have given a window into the rich interplay between geometry and mechanics in thin solids \cite{King12,Vella15}, including connections to pattern formation in liquid crystals %and uncovered fundamental mechanisms for pattern formation with analogs to liquid crystals 
\cite{Aharoni17,Tovkach20,Tobasco20}. %Huang10
Current understanding in this area has been driven primarily by studies on planar sheets \cite{Bico18,Paulsen19}. 
%We show that intrinsically curved shells give rise to qualitatively different behaviors under geometric confinement, enabling the targeted sculpting of liquid surfaces into desired shapes. 
Do thin polymer shells exhibit qualitatively different behaviors from planar sheets, or is the response dictated primarily by the \textit{difference} in curvature between the film and the interface, as suggested by recent work \cite{Taffetani17,Bense20}? % could cite Tobasco19?
More generally, can shells offer new ways to control fluid interfaces, beyond what is possible with planar sheets? 

Here we study the deformations of ultrathin axisymmetric shells on curved liquid interfaces using experiment and theory. 
Surprisingly, we find that over a wide range of parameters, the underlying liquid simply takes on the intrinsic shape of the shell. 
This behavior is distinct from that of planar films, which are inevitably deformed by a curved liquid interface~\cite{King12,Yao13}. The ability to ``sculpt'' a liquid with a polymer shell offers a novel route to controlling the optical properties of an interface.

%\begin{figure}[tb]
%\includegraphics[width=0.94\textwidth]{schematic.pdf}
%\caption{Schematic of an ultrathin spherical cap coming into contact with a liquid interface with a generally different curvature. The interface and shell deform into a new shape, $z(r)$.
%}
%\label{fig:schematic}
%\end{figure}

We form spherical polystyrene shells of Young's modulus $E = 3.4$ GPa and thickness $30<t<631~\si{\nano m}$ by spin coating onto optical lenses with radius of curvature $7< R < 500~\si{\milli m}$~\cite{SM}. 
A circular domain of radius $1.8<W<11.4~\si{\milli m}$ is then cut and delivered to a flat air-water interface with surface tension $\gamma=72~\si{\milli\newton/m}$. 
The mechanical properties of the shell are set by its stretching and bending moduli, $Y=Et$ and $B=Et^3/[12(1-\nu^2)]$ respectively, and its Poisson ratio $\nu=0.34$. 
Our parameters place us in the high bendability regime $\epsilon^{-1}=\gamma W^2/B>10^3$~\cite{King12}: our films buckle under minute compression. 
%One might expect such a thin, floppy shell to be incapable of imposing its shape on a liquid interface, given the near-zero elastic cost of bending. 
%Instead, we show how the mechanical integrity of an ultrathin shell 
As we will show, their ability to impose their shape on a liquid is rooted in the high cost of stretching, analogous to the rigidity of a stiff mylar balloon rather than the geometric rigidity of shells that underlies the strength of architectural domes \cite{Vella12,Lazarus12}. 

\begin{figure*}
\includegraphics[width=0.95\textwidth]{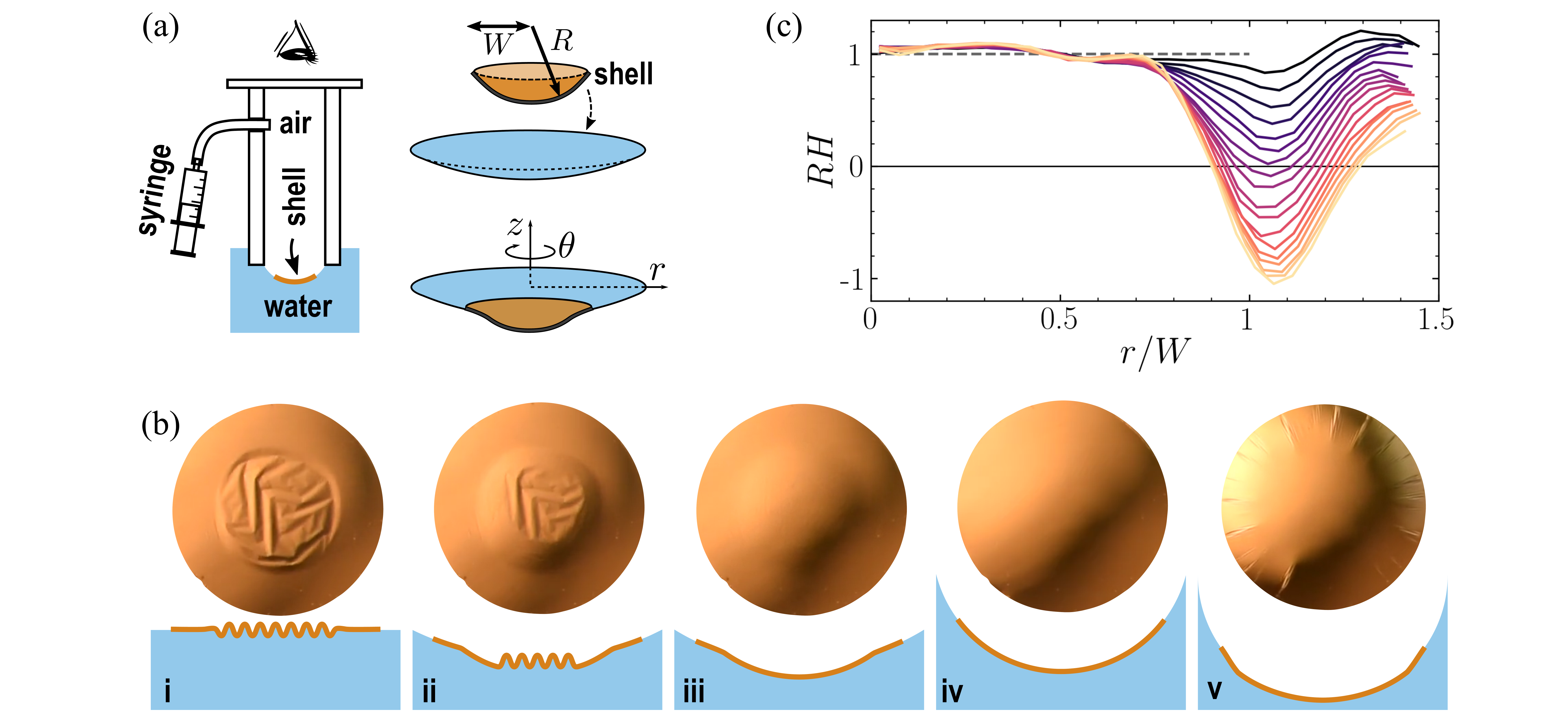}
\caption{\textbf{Sculpting a curved liquid interface.}
(a) Experimental schematic. Adjusting the air pressure in the tube %results in 
leads to different shell configurations.
(b) Top-view images of a deformed shell with $t=60$ nm, $W=2.1$ mm, and $R=13.9$ mm, with side-view schematics. Pressure increases from left to right: \textit{i.}\ a flat interface recovers the behavior in Fig.~\ref{fig:flat}; \textit{ii.}\ part of the wrinkled region ``inflates'' to its rest curvature but a wrinkled core remains; \textit{iii.}\ the entire wrinkled region is inflated; \textit{iv.}\ the interface curvature matches the shell curvature; \textit{v.}\ radial wrinkles grow from the outer edge of the shell. 
(c) %Interface curvature in the radial direction 
Mean interfacial curvature $H$ (non-dimensionalized by the intrinsic curvature of the shell $R^{-1}$) versus $r/W$, for a shell with $t=123$~nm, $W=2.2$~mm, and $R=13.9$~mm. 
The data span from stage \textit{iii.}~(bottom yellow curve) to stage \textit{iv.}~(top curve). 
%The curvature data are non-dimensionalized by the intrinsic curvature of the shell, $R^{-1}$. 
The center of the shell maintains a constant curvature that is close to its intrinsic value (dashed line: $H=R^{-1}$). 
%(d) Interfacial curvature at the center of the inflated shell, $z''(0)$, over a variety of shell curvatures and thicknesses.  
%(Each point is a single measurement; repeated symbols are from the same shell at different pressures.) 
%The values are close to the intrinsic shell curvatures, $R^{-1}$ (dashed line). 
%(e) \textit{Top:} Schematic of an experiment where we place a shell with $W=3.1$ mm and $R = 7.0$ mm on the bottom of a 25 $\mu$L droplet of fluorinated oil ($\rho=1,860$ kg/m$^3$, $\gamma=16$ mN/m) floating on a flat water bath. \textit{Bottom:} The deformed droplet brings part of an image below it into focus. (Detail from \textit{The Garden of Earthly Delights} by Hieronymus Bosch.) Scale: $2$ mm. 
} 
\label{fig:curved}
\end{figure*}

In our experiments, we capture the floating shell with a tube as drawn in Fig.~\ref{fig:curved}(a), so that the interface curvature can be varied continuously by injecting air with a syringe. 
In the top-view images in Fig.~\ref{fig:curved}(b), we observe a central wrinkled ``core'' that shrinks as the interfacial curvature increases. 
We can identify two regions with different curvatures in panels (ii) and (iii), a central core and an outer rim; the core has roughly the same size as the wrinkled region in panel (i).
This distinction disappears in panel (iv), where the curvature seems uniform.
In panel (v), radial wrinkles appear at the edge of the sheet, similar to those observed when a flat sheet is placed on a curved interface~\cite{King12}, suggesting that the liquid interface is more curved than the rest shape of the shell.

%\begin{itemize}
%\item two regions with different curvatures in (ii) and (iii).
%\item the transition between the two regions is roughly at the transition between wrinkled and unwrinkled when the interface is flat.
%\end{itemize}
%At higher pressure where the liquid is more curved than the rest shape of the shell, wrinkles appear at the edge of the film~\cite{King12}.

To quantify the interface shape %throughout 
during this process, we view a checkerboard pattern through the interface; tracking the optical distortion of the pattern allows us to deduce the height profile of the interface using a synthetic Schlieren technique~\cite{Moisy2009, SM, Demery21}. 
Figure~\ref{fig:curved}(c) shows the measured mean curvature $H$ (averaged azimuthally and non-dimensionalized by $R^{-1}$) versus the fractional distance to the center, $r/W$. 
The data are from a range of pressures where no wrinkles are observed [panels (iii) and (iv) in Fig.~\ref{fig:curved}(b)]. 
%As we increase the pressure, the curvature in the center of the shell remains approximately constant %continuously from the bottom yellow curve to the top black curve, 
%and close to the intrinsic curvature of the shell (dashed gray line). 
As we increase the pressure, %we notice that 
the curvature in the center of the shell remains approximately constant and close to the intrinsic curvature of the shell. 
These observations herald the existence of a regime where the shell sculpts the fluid into its rest shape.

\textit{Model.---}
%The mechanical properties of the shell are set by its stretching and bending modulii, $Y=Et$ and $B=Et^3/[12(1-\nu^2)]$ respectively, and the Poisson ratio $\nu$.
The rest shape of the shell is described by an axisymmetric height function $h(r)=r^2/(2R)$, for $0\leq r\leq W$; our shells have small slope, $W\ll R$.
The shell is placed at the interface of a liquid with density $\rho$, and a pressure drop $P_0$ is imposed across the interface at the edge of the shell, setting the curvature of the interface through the Laplace law.

The stresses in the radial and azimuthal directions, $\srr$ and $\stt$, and the height $z$ follow the F\"oppl-von K\'arm\'an equations, which read in polar coordinates~\cite{King12, SM}
\begin{align}
	\partial_r(r\srr)&=\stt,\label{eq:inplane_r}\\
	\partial_r (r\stt)&= \srr+\frac{Y}{2}\left(h'^2-z'^2\right),\label{eq:compatibility}\\
	z''\srr + \frac{z'}{r}\stt & = P_0+\rho g z\ ,\label{eq:vertical_force_bal}
\end{align}
where $g$ is the gravitational acceleration.
The first equation is the in-plane force balance in the radial direction.
The second equation is a compatibility condition, which highlights the role of the mismatch between the rest shape and the actual shape of the sheet as a source of stress.
%(Note that the sign of the rest shape of the shell is unimportant.)
The third equation is the vertical force balance, where we have discarded the bending contribution.
These equations must be supplemented with boundary conditions, provided at $r=0$ by the smoothness of the shape, $z'(0)=0$, the continuity of displacement, $\srr(0)=\stt(0)$; and at $r=W$ by the radial force balance, $\srr(W)=\gamma$, and the convention $z(W)=0$.

% so that the in-plane stress in any direction can only be positive or zero.
We use tension field theory to predict the shape of our shells~\cite{King12, Mansfield2005}: we impose that the stress field in any direction is positive or zero.
A vanishing stress means that compression is released by small scale features such as wrinkles; we do not describe such features and describe instead the gross shape of the sheet through the height function $z(r)$~\cite{Paulsen15}.

%, justifying the use of tension field theory. 
%We use tension field theory to predict the shape of the ultrathin shell with negligible resistance to bending~\cite{King12, Mansfield2005}. 
%Such films buckle under minute compression so that the in-plane stress in any direction can only be positive or zero.
%Exploiting the cylindrical symmetry of our system, we write down the F\"oppl-von K\'arm\'an (FvK) equations in polar coordinates~\cite{King12}:
%\begin{align}
%	\partial_r(r\srr)&=\stt,\label{eq:inplane_r}\\
%	\partial_r (r\stt)&= \srr+\frac{Y}{2}\left(h'^2-z'^2\right),\label{eq:compatibility}\\
%	z''\srr + \frac{z'}{r}\stt & = -P_0+\rho g z\ ,\label{eq:vertical_force_bal}
%\end{align}
%where $g$ is the gravitational acceleration and $P_0$ is the pressure drop across the interface at the edge of the sheet where we set $z(W)=0$.
%The first equation is the in-plane force balance in the radial direction.
%The second equation is a compatibility condition, which highlights the role of the mismatch between the rest shape and the actual shape of the sheet as a source of stress.
%(Note that the sign of the rest shape of the shell is unimportant.)
%The third equation is the vertical force balance, where we have discarded the bending contribution. % in the vertical force balance.
%These equations must be supplemented with boundary conditions, provided at $r=0$ by the smoothness of the shape, $z'(0)=0$, the continuity of displacement, $\srr(0)=\stt(0)$; and at $r=W$ by the radial force balance, $\srr(W)=\gamma$, and the convention $z(W)=0$.

\begin{figure*}[t]
\includegraphics[width=0.99\textwidth]{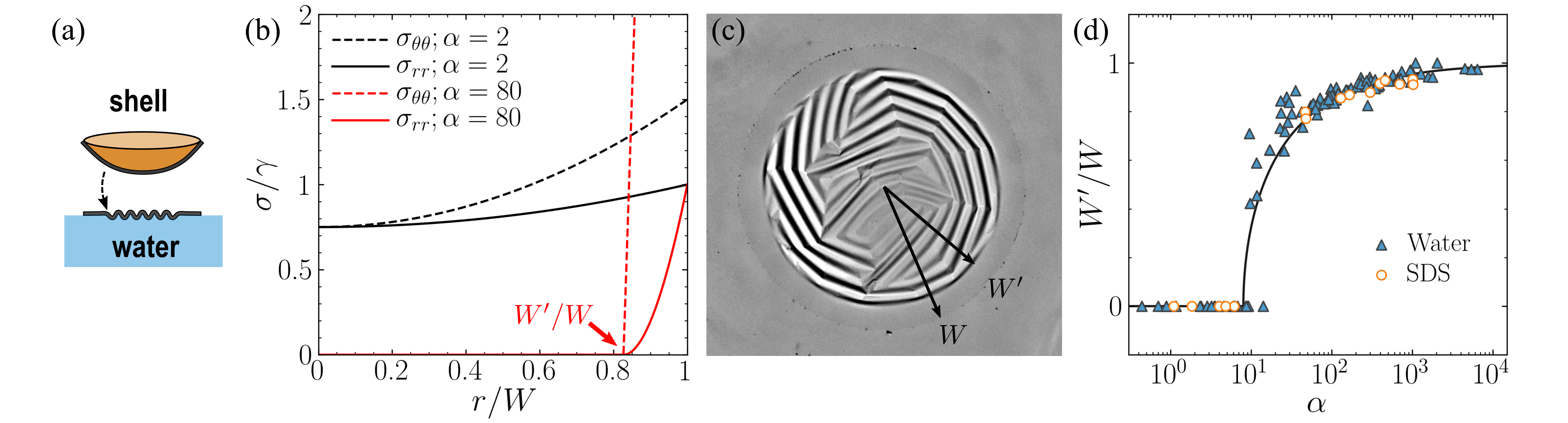}
\caption{\textbf{Stretching and wrinkling on a flat liquid interface.} (a) Experimental schematic. %Schematic of placing a thin curved shell on a flat liquid bath.
(b) Radial (solid line) and azimuthal (dashed line) stress in the sheet from the analytic solution.
(c) Top-view image of an ultrathin shell ($t=112$ nm, $W=6.6$ mm, $R=51.5$ mm) conforming to a flat liquid interface by forming a wrinkled core %region 
of radius $W'$ and an unwrinkled rim. Background subtracted for clarity.
(d) $W'/W$ versus $\alpha$ for shells with $30<t<631$ nm, $13.8<R<500$ mm, and $2.2<W<11.4$ mm on water ($\gamma= 72$ mN/m, filled symbols) or an aqueous solution of sodium dodecyl sulfate ($\gamma = 36$ mN/m, open symbols). Solid line: Theory with no free parameters [Eq.~(\ref{eq:W_prime})].
}
\label{fig:flat}
\end{figure*}

\textit{Flat interface.---} We first consider the situation where no pressure drop is imposed across the interface, $P_0=0$ [Fig.~\ref{fig:flat}(a)]. 
In this case the sheet remains flat, as $z=0$ solves the vertical force balance [Eq.~(\ref{eq:vertical_force_bal})]. 
Then, the solution to Eqs.~(\ref{eq:inplane_r}, \ref{eq:compatibility})  
depends only on the dimensionless \emph{confinement parameter}~\cite{King12, SM}
\begin{equation}\label{eq:confinement}
\alpha = \frac{YW^2}{2\gamma R^2},
\end{equation}
which compares the tension applied at the edge, $\gamma$, to the stress 
that is required to flatten the shell, $YW^2/R^2$.
There is a critical value of the confinement, $\alpha_c=8$, below which the stresses remain positive over the whole sheet [Fig.~\ref{fig:flat}(b), black lines].

On the contrary, above the critical value, the solution 
%to Eqs.~(\ref{eq:inplane_r}, \ref{eq:compatibility})
should vanish in a circular region around the center of the sheet, indicating the appearance of small-scale features [Fig.~\ref{fig:flat}(c)]. 
%Under tension field theory, when the stress in direction $i$ becomes negative, the force balance in direction $i$ has to be replaced by the equation $\sigma_{ii}=0$.
Inspection of Eqs.~(\ref{eq:inplane_r}, \ref{eq:compatibility}) shows that the stress vanishes in the same region in the two directions: $\srr=0$ and $\stt=0$ for $r<W'$, so that the boundary condition at $r=0$ has to be replaced by the condition $\srr(W')=0$.
Solving the force balance equations with the new boundary condition for $W'<r<W$ provides the stress field in the sheet [Fig.~\ref{fig:flat}(b), red lines], and the value of $W'$:
\begin{equation}
\label{eq:W_prime}
\frac{W'}{W}=\sqrt{1-\sqrt{\frac{\alpha_c}{\alpha}}}.
\end{equation}
We thus predict a central wrinkled region 
%region in the center of the shell
whenever $\alpha \geq \alpha_c$, having a size $W'$ that grows continuously with $\alpha$, reaching $W'=W$ in the limit $\alpha\to\infty$ [Fig.~\ref{fig:flat}(d), solid line]. 
A similar result was obtained for %the size of 
a neutral scarred zone in a crystalline domain bound to a sphere \cite{Azadi16}.
The sheet remains unwrinkled in a rim of width $L=W-W'$, which becomes independent of %the radius of 
the sheet size
at large confinement: 
\begin{equation}\label{eq:rim}
L\sim W\sqrt{\frac{2}{\alpha}}=2R\sqrt{\frac{\gamma}{Y}}.
\end{equation}
%The shell forms disordered wrinkles in the central region surrounded by a flat rim, 
Our experiments on a flat bath support these predictions. 
Figure \ref{fig:flat}(c) shows a circular region of disordered wrinkles surrounded by an unwrinkled rim. 
The radius $W'$ of the wrinkled region is plotted in Fig.~\ref{fig:flat}(d) as a function of the confinement $\alpha$. 
The data fall onto Eq.~(\ref{eq:W_prime}) over 4 orders of magnitude in $\alpha$ with no free parameters. 
We also find good agreement at a second value of surface tension. % [open symbols in Fig.~\ref{fig:flat}(d)].

\textit{Curved interface.---}
We turn to the situation where a pressure difference is imposed across the interface. 
%We first consider the case of a spherical shell before generalizing to an arbitrary shape. 
Once again there are solutions to Eqs.~(\ref{eq:inplane_r}-\ref{eq:vertical_force_bal}) with a wrinkled core or without one. 
If there is a wrinkled core with radius $W'$, then the hydrostatic pressure vanishes there: $z(r)=-P_0/(\rho g)$ for $0\leq r\leq W'$,  
consistent with the vertical force balance (\ref{eq:vertical_force_bal}) in the absence of stress. 
This sets the boundary condition at $r=W'$. 
In the unwrinkled portion, we integrate Eqs.~(\ref{eq:inplane_r}-\ref{eq:vertical_force_bal}) %are integrated 
numerically using the boundary value problem solver \emph{integrate.solve\_bvp} implemented in SciPy.
%, with the confinement $\alpha=YW^2/(2\gamma R^2)$ and the pressure $P_0$ as the control parameters. 

\begin{centering}
\begin{figure}[tb]
\includegraphics[width=0.9\textwidth]{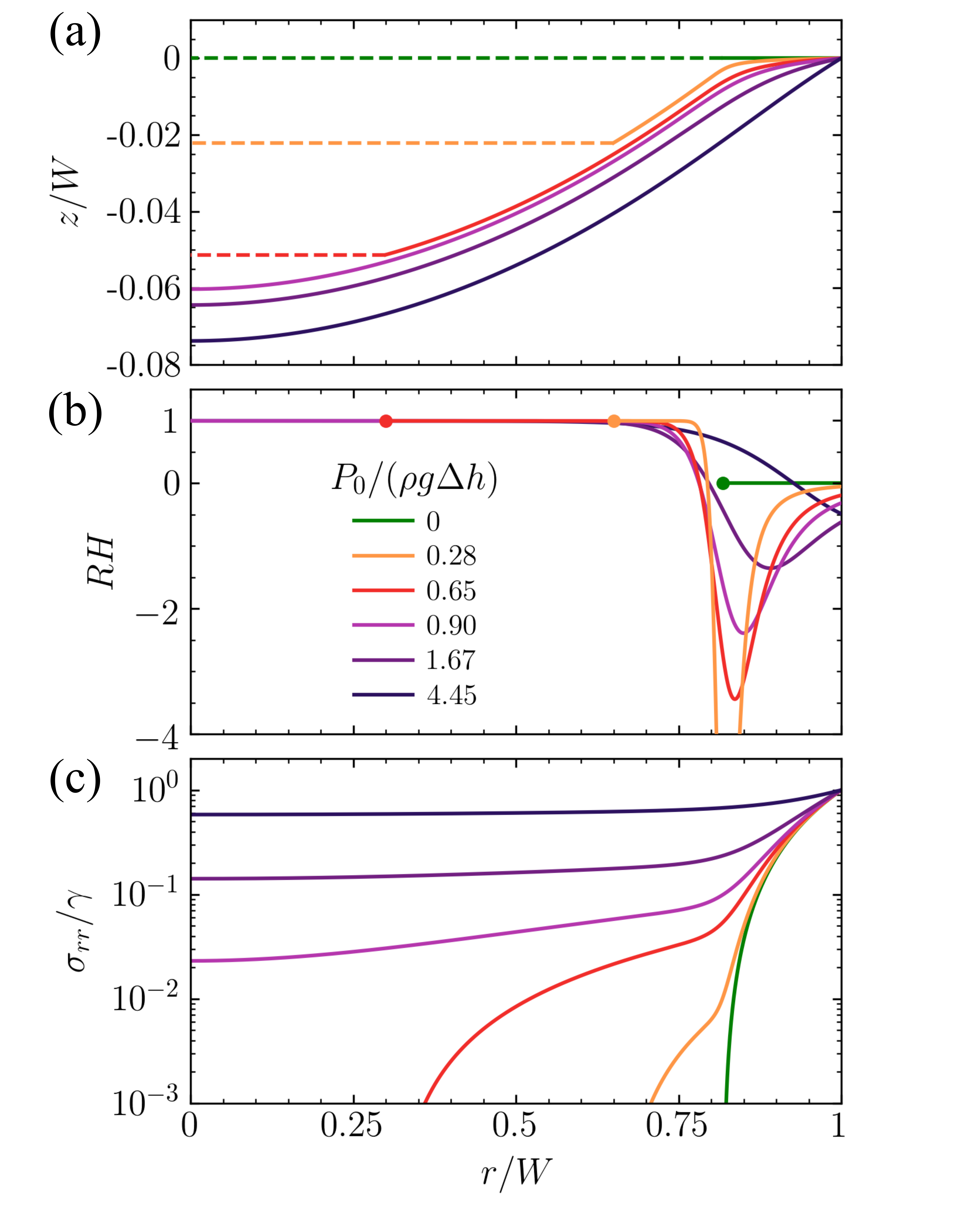}
\caption{\textbf{Numerical solution on a curved interface with $\alpha>\alpha_c$.}
We use $R/W=6.3$, $\gamma/Y=1.7\times 10^{-4}$, $W^2\rho g/Y=1.1\times 10^{-4}$ ($\alpha=73$).
(a) Profiles of shells at different pressures.  Dashed lines indicate wrinkled regions.  The (partially) inflated regions nearly match the initial shape.  
(b) Curvature versus normalized radial position $r/W$. %Filled 
Dots indicate boundaries between wrinkled and smooth regions.
A curvature $RH=1$ corresponds to the initial shape of the shell.
(c) Radial stress versus normalized radial position $r/W$.
Green curves: %are the 
Analytic solution for a flat %liquid 
interface ($P_0=0$). 
}
\label{fig:theo}
\end{figure}
\end{centering}

Figure~\ref{fig:theo} shows the numerical results %for 
corresponding to the sheet %used 
in Fig.~\ref{fig:curved}(c), which has a confinement $\alpha=73\gg\alpha_c$; we plot the profile of the sheet $z(r)$, its mean curvature $H(r)=[z''(r)+z'(r)/r]/2$ and the radial stress field $\srr(r)$ for different values of the pressure $P_0$.
The top curve of Fig.~\ref{fig:theo}(a) shows that at zero pressure, there is a wrinkled zone in the center and an unwrinkled rim at the edge; this is simply the flat interface case of Fig.~\ref{fig:flat}. %, as we saw in the experiments on a flat interface. 
For small positive pressure, the wrinkled region ``inflates'' to the height $z^*=-P_0/(\rho g)$ with wrinkles persisting in the %central part 
center where $z=z^*$. 
As in the flat case, the radial stress falls to $0$ at the edge of the wrinkled region  [Fig.~\ref{fig:theo}(c)]. 
Remarkably, between the wrinkled region and the outer rim, the profile of the sheet is very close to its shape at rest: $RH\simeq 1$ [Fig.~\ref{fig:theo}(b)].
If the pressure is large enough, the sheet deploys completely: the wrinkles in the center are gone and the sheet is under tension everywhere. 
We find that the size of this ``inflatable'' region is close to %the size 
that of the wrinkled region when the same shell is %placed 
on a flat bath, %interface:
so that 
the size of the rim on a curved interface is also given by Eq.~(\ref{eq:rim}).
%Thus, Eq.~(\ref{eq:rim}) for the size of the rim  also applies on a curved interface.
This phenomenology matches the experimental observations (Fig.~\ref{fig:curved}, see~\cite{SM} for a quantitative comparison).
The behavior is very different at small confinement, where there would be no wrinkles on a flat interface: in this case the shell departs significantly from its rest shape on a curved interface~\cite{SM}.

A key quantity is the minimum pressure $P_\textrm{c}$ needed to inflate the shell completely.
It can be estimated as $P_\textrm{grav}=\rho g\Delta h$, where $\Delta h=W^2/(2R)$ is the initial height of the shell.
%=|h(0)-h(W)|
For the parameters in Fig.~\ref{fig:theo}, we find $P_\textrm{grav} \simeq 0.20 P_\textrm{Lap}$, where $P_\textrm{Lap} = 2\gamma/R$ is the Laplace pressure required to create a liquid interface of the same curvature. 
This estimate is an upper bound due to the flattened rim; the pressure needed to inflate the sheet in our numerical solution is $0.16 P_\textrm{Lap}$. 
%Thus, the shell imposes its shape over a wide range of pressures (\textit{i.e.}, from this value up to $P_\textrm{Lap}$).

\begin{figure}
\begin{center}
\includegraphics[width=0.92\textwidth]{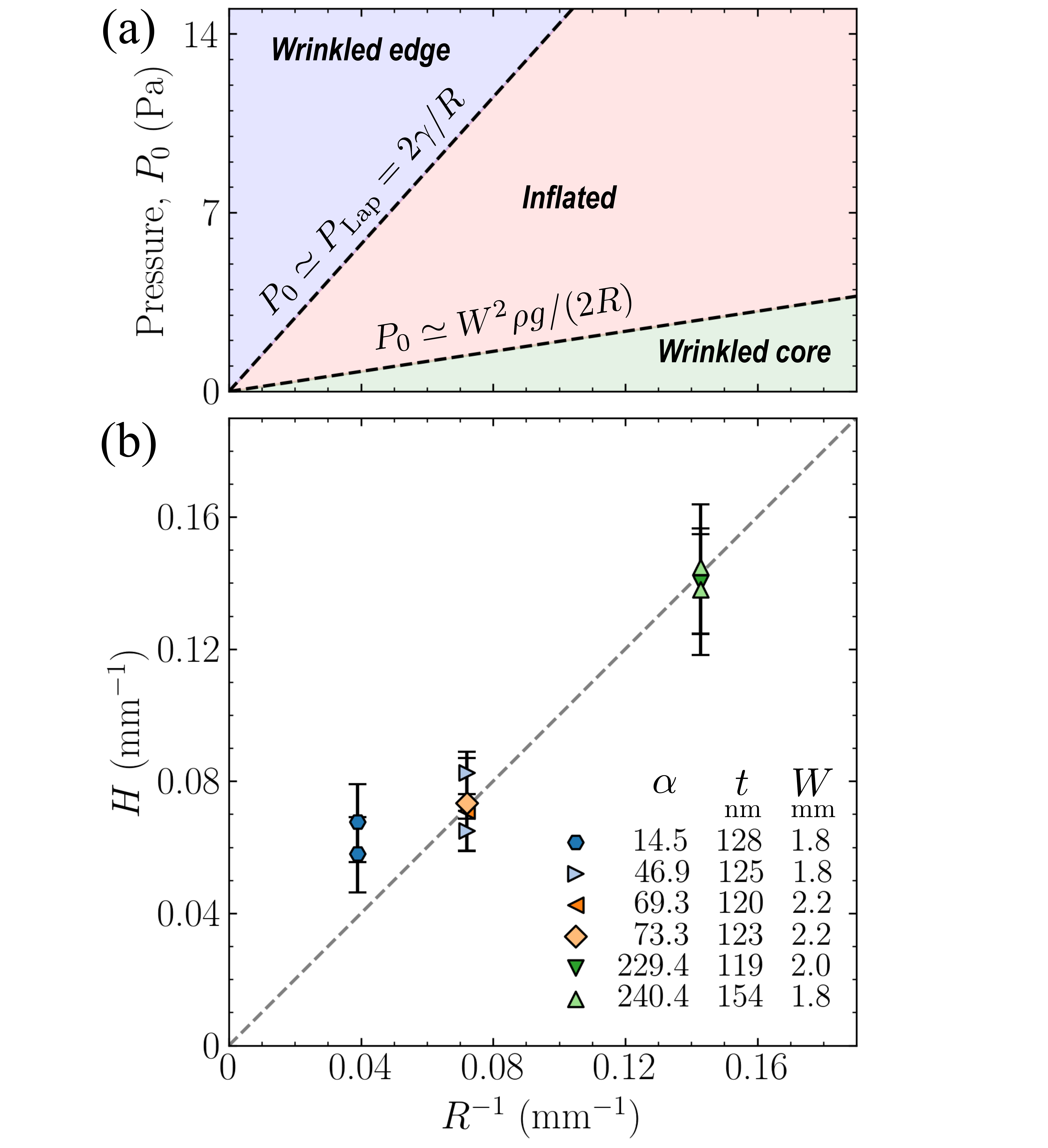}
\end{center}
\caption{
\textbf{Inflated regime.}
(a) Phase diagram in the inextensible limit. 
%at large confinement. Dashed lines show the phase boundaries 
(b) %Interfacial 
Mean curvature at the center of the %inflated 
shell %within 
in the ``inflated'' regime, for shells with a large confinement. 
The values are close to the intrinsic shell curvatures, $R^{-1}$ (dashed line), for a variety of shell curvatures and thicknesses. 
%Each point is a single measurement; 
Repeated symbols are from the same shell at different pressures. 
}
\label{fig:phase_diag}
\end{figure}

In the inextensible limit $Y\to \infty$,
%(while keeping a negligible bending modulus), 
the rim disappears and the sheet is perfectly inflated for %$P\in[P_\textrm{inf},P_\textrm{Lap}]$.
$P_\textrm{grav} < P < P_\textrm{Lap}$.
This range of pressures is shown in Fig.~\ref{fig:phase_diag}(a) as a function of the curvature of the shell; there is a wrinkled core for $P<P_\textrm{grav}$ and a wrinkled edge for $P>P_\textrm{Lap}$. 
For a finite stretching modulus, the range of the ``inflated region'' increases while the size of the inflated core shrinks [Eq.~(\ref{eq:W_prime})]. 
%Previous e
Experiments with a flat sheet by King et al.~\cite{King12} correspond to the vertical axis %on the phase diagram 
of Fig.~\ref{fig:phase_diag}(a)
at zero curvature; there only the ``wrinkled edge'' region is accessible. 
We validate this theoretical picture by entering the inflated regime in 5 additional experiments spanning a range of curvatures and thickness, all at large confinement. 
Each shell inflates to its original shape: the measured curvature in the center of the shell is in agreement with the intrinsic shell curvature [Fig.~\ref{fig:phase_diag}(b)].

\textit{Discussion.---}
We have shown how a thin interfacial shell with vanishing bending rigidity behaves qualitatively differently than a planar film. 
Namely, a shell may impose its own shape on an interface over a range of pressures, offering a straightforward method to control the equilibrium shape of a fluid. 
%We have shown how a thin shell with vanishing bending rigidity may nevertheless impose its own shape on a liquid interface. 
One advantage of this self-inflating regime is that the deployed shape is robust to perturbations in pressure, unlike a bare liquid interface where the curvature varies continuously with the Laplace pressure. 
This property could be useful for optical applications, and it may be achieved with little intervention, which we demonstrate by inflating a shell using an oil droplet floating on water~\cite{SM}. 

Although we focused on spheres, our analysis can be generalized to any axisymmetric shell.
%As an example, we solved the equations for a conical shell numerically at large confinement and observed the same qualitative behavior~\cite{SM}: the region that is wrinkled when $P_0=0$ inflates to its rest shape when the pressure is increased.
When the stretched rim is narrow, its size should depend only on the slope of the shell at the edge, $h'(W)$, since this is the sole aspect of the shape that appears explicitly in the force balance [Eqs.~(\ref{eq:inplane_r}-\ref{eq:vertical_force_bal})].
%We may then derive an expression for the rim size at large confinement:
Writing Eq.~(\ref{eq:rim}) using the slope $h'(W)=W/R$, we find $L\simeq 2W\sqrt{\gamma/Y}/h'(W)$.
%\begin{equation}\label{eq:rim2}
%L\simeq \frac{2W}{h'(W)}\sqrt{\frac{\gamma}{Y}}.
%\end{equation}
%Comparing this expression to Eq.~(\ref{eq:rim}), we obtain an expression for the ``effective confinement'', $\alpha=Yh'(W)^2/(2\gamma)$.
%The tension applied at the edge of the shell is damped over this rim width, leaving the center of the sheet at rest. 
%At small confinement, the tension propagates all the way to the center of the sheet. 
This generalization is supported by a detailed analysis of a conical shell on a curved interface~\cite{SM}. 
Moreover, our numerical results for a cone %on an %curved 
%interface 
show that the region %of the cone 
that is wrinkled for $P_0=0$ corresponds to the region that inflates to its rest shape %when the pressure is increased
at sufficient pressure, just as it does for a spherical shell~\cite{SM}.
%As an example, we solved the equations for a conical shell numerically at large confinement and observed the same qualitative behavior~\cite{SM}: the region that is wrinkled when $P_0=0$ inflates to its rest shape when the pressure is increased.

Not all axisymmetric shells %will 
inflate to their rest shape. % when a pressure drop is applied across the interface. 
The question of which shapes are maintained upon inflation dates back to the optimization of parachutes by Taylor~\cite{Taylor63}. Since then, closed surfaces have received the most attention~\cite{Paulsen94, Pak10}.
%In particular, 
Recently, Gorkavyy %gave the 
reported a condition for an axisymmetric shell to retain its shape upon inflation~\cite{Gorkavyy2010}, although 
%However, 
these calculations are  
for a uniform pressure drop across the shell; the condition in the presence of a pressure gradient is as yet unknown.
Whatever this condition may be, our work suggests that it is satisfied for a sphere and a cone.

\begin{acknowledgements}
We are grateful to G.~M.~Grason and F.~Montel for useful discussions, G.~C.~Leggat for help with an early version of the experiment, and S.~Prasch in the Syracuse University Glass Shop for assisting with the tubes. 
We thank P.~Damman and D.~Vella for useful comments on the manuscript. 
This work was supported by NSF Grants No.~DMR-CAREER-1654102 (Y.T. and J.D.P.) and No.~REU DMR-1460784 (A.R.H.). 
J.D.P. gratefully acknowledges support from the ESPCI Paris Total Chair. 
\end{acknowledgements}

%\bibliography{bibsheets-shortTaylor}

\end{document}